# Band structure of SrFeAsF and CaFeAsF as parent phases for a new group of oxygen-free FeAs superconductors.

I.R. Shein,* and A.L. Ivanovskii

*Institute of Solid State Chemistry, Ural Branch of the Russian Academy of Sciences, Ekaterinburg, GSP-145, 620041, Russia*

**Abstract**

By means of first-principle FLAPW-GGA calculations, we have investigated the electronic properties of the newly discovered layered quaternary systems SrFeAsF and CaFeAsF as parent phases for a new group of oxygen-free FeAs superconductors. The electronic bands, density of states, Fermi surfaces, atomic charges, together with Sommerfeld coefficients $\gamma$ and molar Pauli paramagnetic susceptibility $\chi$ have been evaluated and discussed in comparison with oxyarsenide LaFeAsO - a parent phase for a new class of high-temperature ($T_C \sim 26\text{-}56K$) oxygen-containing FeAs superconductors. Similarity of our data for SrFeAsF and CaFeAsF with the band structure of oxygen-containing FeAs superconducting materials may be considered as theoretical background specifying the possibility of superconductivity in these oxygen-free systems.

\_\_\_\_\_\_\_\_\_\_\_\_\_\_\_\_\_\_\_\_\_\_\_\_\_\_\_\_\_\_\_\_

\* *E-mail address*: shein@ihim.uran.ru

# 1. Introduction

The recent discovery [1-7] of new FeAs superconductors (SCs) with the transition temperatures to $T_C \sim 56K$ has attracted great interest in these materials since it is the first class of copper-free systems which exhibit superconductivity at high temperatures. This surprising discovery has stimulated much activity in search for new related superconducting materials, and element substitution seems to be an efficient way to produce new SCs of this family. However until now most of the research groups have focused on the most numerous group of these SCs which are based on layered quaternary oxyarsenides $Ln$FeAsO, (where $Ln$ = La, Ce … .Gd, Tb, Dy). These phases adopt a quasi-two-dimensional ZrCuSiAs-like crystal structure with alternating [FeAs] and [$Ln$O] molecular layers.

Quite recently, new oxygen-free ZrCuSiAs-like phases $A$FeAsF (where $Ln$ = Sr, Ca and Eu) have been synthesized [8-11] and some of their properties have been determined. In particular, strong anomalies in specific heat, electrical resistance and magnetic susceptibility (which have been interpreted as spin density wave (SDW) anomalies) at about 173K for SrFeAsF, 118K for CaFeAsF and 153K for EuFeAsF were observed [8-11]. Since similar SDW anomalies are known to be an important prerequisite for high-$T_C$ superconductivity in oxygen-containing FeAs materials, it was assumed [8] that $A$FeAsF can serve also as parent phases for a new oxygen -free group of FeAs SCs. Moreover, critical transitions at $T_C \sim 4K$ in the Co-substituted SrFeAsF (SrFe$_{0.875}$Co$_{0.125}$AsF) [9] and at $T_C \sim 31K$ in the La-substituted SrFeAsF (Sr$_{0.8}$La$_{0.2}$FeAsF) were reported [11].

In this Communication, using the first principles FLAPW method within the generalized gradient approximation (GGA) for the exchange-correlation potential we explore for the first time the electronic properties of two layered arsenides SrFeAsF and CaFeAsF as parent phases for a new group of oxygen-free FeAs superconductors.

## 2. Computational details

Our calculations were carried out by means of the full-potential method with mixed basis APW+lo (FLAPW) implemented in the WIEN2k suite of programs [12]. The generalized gradient approximation (GGA) to exchange-correlation potential in the PBE form [13] was used. The plane-wave expansion was taken up to $R_{MT} \times K_{MAX}$ equal to 7, and the $k$ sampling with $10 \times 10 \times 4$ $k$-points in the Brillouin zone was used.

The calculations were performed with full-lattice optimizations including the so-called internal parameters $z_{Sr(Ca)}$ and $z_{As}$. The self-consistent calculations were considered to be converged when the difference in the total energy of the crystal did not exceed 0.1 mRy and the difference in the total electronic charge did not exceed 0.001 $e$ as calculated at consecutive steps.

The analysis of the hybridization effects was performed using the densities of states (DOS), which were obtained by a modified tetrahedron method [14]. The ionic bonding was studied with the use of Bader [15] analysis.

## 3. Results and discussion

*3.1. Structural properties.*

As the first step, the total energy ($E_{tot}$) *versus* cell volume calculations were carried out to determine the equilibrium structural parameters for SrFeAsF and CaFeAsF; the calculated values are presented in Table 1. These data are in reasonable agreement with the available experiments [8-11]. As can be seen, when going from SrFeAsF to CaFeAsF the inter-layer distances decrease and simultaneous compression of the layers in *xy* planes takes place. This result can be easily explained taking into account the values of the ionic radii: R($Sr^{2+}$) = 1.12 Å > R($Ca^{2+}$) = 0.99 Å. However some *anisotropic deformation* of the crystal structure as going from SrFeAsF to CaFeAsF takes place: according to our calculations, the change of the parameter *c* (at about 0.5 Å) is higher than for the parameter *a* (at about 0.1 Å).

*3.2. Electronic properties.*

Figure 1 shows the band structures of SrFeAsF to CaFeAsF as calculated along the high-symmetry *k* lines. The *E(k)* curves in the high-symmetry directions in the Brillouin zone (BZ) demonstrate evident similarities in the energy bands for these isoelectronic and isostructural phases. Let us discuss the main peculiarities of the electronic structure of SrFeAsF to CaFeAsF using their densities of states (DOSs) as depicted in Figure 2.

As can be seen, the quasi-core As 4*s* states are located from -12.4 eV to -10.2 eV below the Fermi level ($E_F$) and are separated from the valence states by a gap. The valence band (VB) extends from -6.8 eV (for SrFeAsF) and from -7.3 eV (for CaFeAsF) up to the Fermi level $E_F$ = 0 eV and includes three main subbands *A-C*, Figure 2. Among them the first subband *A* is formed mainly by F 2*p* states with some admixture of Sr(Ca) *sp* states, whereas the subband *B* is formed predominantly by hybridized As 4*p* and Fe 3*d* states. Finally, the near-Fermi subband *C* contains the main contributions from the Fe 3*d* states. Thus, the preliminary conclusion from our DOSs calculations is that the general bonding mechanism in both phases is not of a "purely" ionic character owing to the mentioned hybridization of valence states, but includes also covalent interactions inside [LaAs] and [Sr(Ca)F] layers, see also below.

The most interesting feature of the band structures of SrFeAsF and CaFeAsF is the behavior of quasi-flat electronic bands along Γ-Z and A-M directions, which is also found for all other tetragonal oxygen-containing FeAs superconductors. The corresponding Fermi surfaces (FS) in the first BZ are shown in Figure 3. Owing to the quasi-two-dimensional electronic structure, the Fermi surfaces are composed of four cylindrical-like sheets, parallel to the $k_z$ direction. Two of them are of a hole-like character and are centered along the Γ - Z high symmetry lines, whereas the other two sheets are of the electronic-like type and are aligned along the A - M direction. Similar FS topology was established also for the oxygen-containing FeAs superconductors.

As electrons near the Fermi surface are involved in the formation of the superconducting state, it is important to figure out their nature. The total and orbital decomposed partial DOSs at the Fermi level, $N(E_F)$, are shown in Table 2. It is seen that for both SrFeAsF and CaFeAsF phases the main contribution to $N(E_F)$ is from the Fe $3d$ states, and $N(E_F)$ increases as going from SrFeAsF to CaFeAsF. The obtained data allow us also to estimate the Sommerfeld constants (γ) and the Pauli paramagnetic susceptibility (χ) for SrFeAsF and CaFeAsF under assumption of the free electron model as $\gamma = (\pi^2/3)N(E_F)k_B^2$ and $\chi = \mu_B^2 N(E_F)$. It is seen from Table 2 that both γ and χ increase when Sr is replaced by Ca.

Note that our results indicate that both SrFeAsF and CaFeAsF are at the border of magnetic instability. For example, the calculated energy difference between their non-magnetic and ferromagnetic states of SrFeAsF is very small (at about 0.001 eV/form.unit). The calculated local magnetic moment of Fe for ferromagnetic state of SrFeAsF is about 0.15 $\mu_B$.

Finally, let us discuss the bonding picture for SrFeAsF and CaFeAsF in greater detail.

*3.3. Chemical bonding.*

To describe the ionic bonding for SrFeAsF and CaFeAsF, it is possible to start with a simple ionic picture, which considers the usual oxidation numbers of atoms: $(Ca,Sr)^{2+}$, $Fe^{2+}$, $As^{3-}$ and $F^{1-}$. Thus, the charge states of the layers are $[Sr(Ca)F]^{1+}$ and $[FeAs]^{1-}$, *i.e.* the charge transfer occurs from $[Sr(Ca)F]^{1+}$ to $[FeAs]^{1-}$ layers. Besides, inside [Sr(Ca)F] and [FeAs] layers, the ionic bonding takes place respectively between Sr(Ca)-F and Fe-As atoms.

To estimate the amount of electrons redistributed between the adjacent [Sr(Ca)F] and [FeAs] layers and between the atoms inside each layer, we carried out Bader analysis. The total charge of an atom (the so-called Bader charge, $Q^B$), as well as the corresponding charges as obtained from the purely ionic model ($Q^i$) and their differences ($\Delta Q = Q^B - Q^i$) are presented in Table 3. These results show that the inter-layer charge transfer is much smaller than it is predicted in the idealized ionic model. Namely, the transfer $\Delta Q([Sr(Ca)F] \rightarrow [FeAs])$ is about 0.31 e and 0.28 for SrFeAsF and CaFeAsF, respectively, and these values are smaller than those for LaFeAsO (0.39 e [16]).

Taking into account the covalent intra-atomic interactions derived from analysis of the site-projected DOS calculations, the common picture of chemical bonding for SrFeAsF and CaFeAsF may be described as follows.

1. Inside [Sr(Ca)F] layers, the ionic Sr(Ca)-F bonds take place with small hybridization of valence states of Sr(Ca)-F;
2. Inside [FeAs] layers, mixed metallic-ionic-covalent Fe-As bonds appear (owing to hybridization of valence states of Fe-Fe and Fe-As atoms and Fe $\rightarrow$ As charge transfer); in addition, inside [FeAs] layers covalent bonds As-As take place (owing to As $4p$ - As $4p$ hybridization);
3. Between the adjacent [Sr(Ca)F] and [FeAs] layers, ionic bonds emerge owing to [Sr(Ca)F] $\rightarrow$ [FeAs] charge transfer, and these bonds are responsible for the cohesive properties of these crystals.

Generally, the bonding in SrFeAsF and CaFeAsF can be classified as a mixture of metallic, ionic and covalent contributions.

## 4. Conclusions

In summary, we have performed FLAPW-GGA calculations to obtain the structural, electronic properties and the chemical bonding picture for two quaternary layered phases SrFeAsF and CaFeAsF as parent phases for a new group of oxygen-free FeAs superconductors.

The comparison of our data for SrFeAsF and CaFeAsF with the band structure of oxygen-containing FeAs SCs reveals close similarity of their electronic properties. These results may be considered as the theoretical background specifying the possibility of superconductivity in these materials. Naturally, further in-depth studies are necessary to understand possible scenarios of doping-induced superconducting coupling mechanisms, as well as the relationships between magnetism and superconductivity for these oxygen-free materials.

Table 1.
The optimized lattice parameters (*a* and *c*, in Å), internal coordinates ($z_{Sr(Ca)}$ and $z_{As}$), some inter-atomic distances (*d*(Fe-As), in Å) and angles (As-Fe-As, in deg.) and cell volumes  (*$V_o$*, in Å$^3$) for tetragonal (space group P4/nmm) layered SrFeAsF and CaFeAsF as compared with available experiments [8-11].

| Phase /parameter | SrFeAsF | CaFeAsF |
|---|---|---|
| *a* | 4.0055 (3.9930 [8]; 3.999 [9]; 4.004 [10]; 4.011 [11]) | 3.9049 |
| *c* | 8.8049 (8.9546 [8]; 8.973 [9]; 8.971 [10]; 8.965 [11]) | 8.3565 |
| *c/a* | 2.1982 (2.2426 [8]; 2.2438 [9]; 2.2405 [10]; 2.2351 [11]$^1$) | 2.1400 |
| $z_{As}$ | 0.6397 (0.6527 [8]) | 0.6525 |
| $z_{Sr,Ca}$ | 0.1665 (0.1598 [8]) | 0.1588 |
| *d*(Fe-As) | 2.350 (2.420 [8]) | 2.331 |
| (As-Fe-As) | 105.89 (108.6 [8]) | 107.38 |
| *$V_o$* | 141.27 (142.77 [8]) | 127.42 |

$^1$ the available experimental data are given in parentheses.

Table 2.
Total $N^{tot}(E_F)$ and partial $N^l(E_F)$ densities of states at the Fermi level (in states/eV·atom$^{-1}$), electronic heat capacity γ (in mJ·K$^{-2}$·mol$^{-1}$) and molar Pauli paramagnetic susceptibility χ (in 10$^{-4}$ emu/mol) of quaternary SrFeAsF and CaFeAsF.

| System /parameter | SrFeAsF | CaFeAsF |
|---|---|---|
| $N^{Fed}(E_F)$ | 1.188 | 1.529 |
| $N^{As}(E_F)$ | 0.039 | 0.040 |
| $N^{tot}(E_F)$ | 1.540 | 1.895 |
| γ | 3.630 | 4.466 |
| χ | 0.496 | 0.610 |

Table 3.
Atomic charges (in e) for SrFeAsF and CaFeAsF and for [FeAs], and [Sr(Ca)O] layers as obtained from a purely ionic model ($Q^i$), Bader analysis ($Q^B$) and their differences ($\Delta Q = Q^B - Q^i$) – in comparison with LaFeAsO [16].

|  |  | Sr(Ca,La) | Fe | As | F(O) | [Sr(Ca,La)F(O)] | [FeAs] |
|---|---|---|---|---|---|---|---|
| SrFeAsF | $Q^i$ | +2 | +2 | -3 | -1 | +1 | -1 |
|  | $Q^B$ | 8.465 | 7.741 | 5.950 | 7.843 | 16.309 | 15.585 |
|  | $\Delta Q$ | 0.465 | 1.741 | -2.050 | -0.157 | 0.309 | -0.309 |
| CaFeAsF | $Q^i$ | +2 | +2 | -3 | -1 | +1 | -1 |
|  | $Q^B$ | 6.411 | 7.734 | 5.986 | 7.869 | 14.280 | 15.603 |
|  | $\Delta Q$ | 0.411 | 1.734 | -2.014 | -0.131 | 0.280 | -0.280 |
| LaFeAsO | $Q^i$ | +3 | +2 | -3 | -2 | +1 | -1 |
|  | $Q^B$ | 9.115 | 7.722 | 5.887 | 7.276 | 16.391 | 14.998 |
|  | $\Delta Q$ | 1.115 | 1.772 | -2.113 | -0.724 | 0.391 | -0.391 |

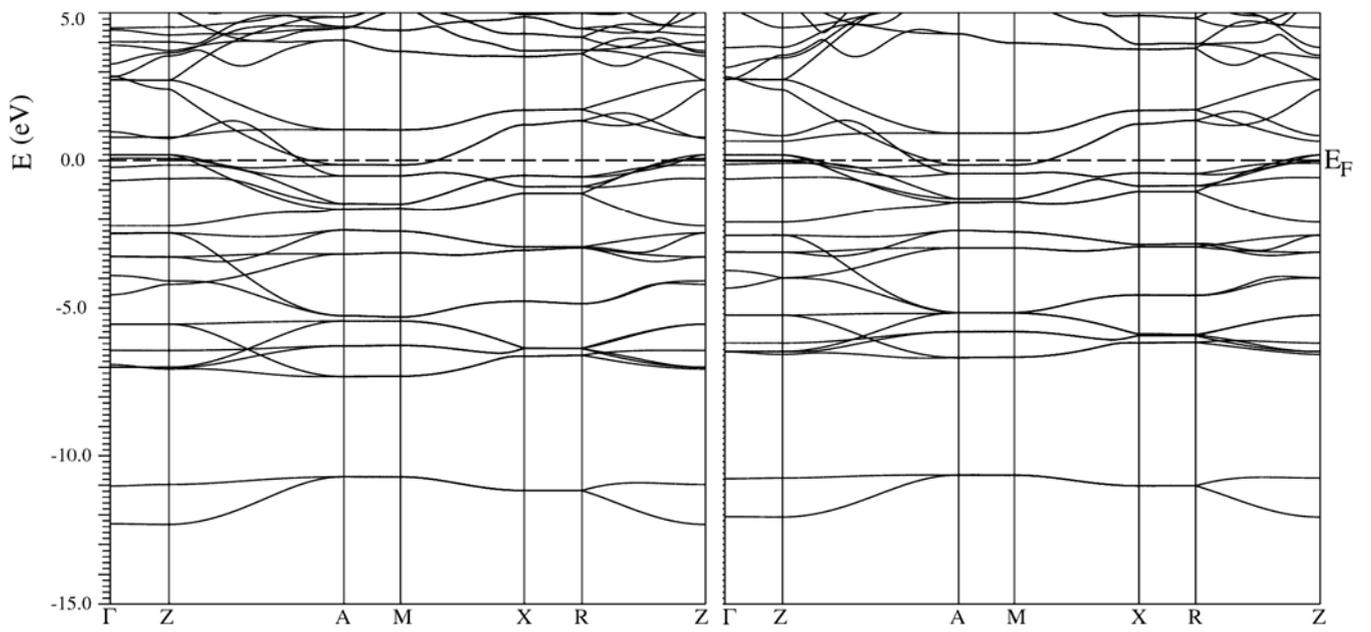

Fig. 1. Electronic band structures of CaFeAsF (*left*) and SrFeAsF (*right*).

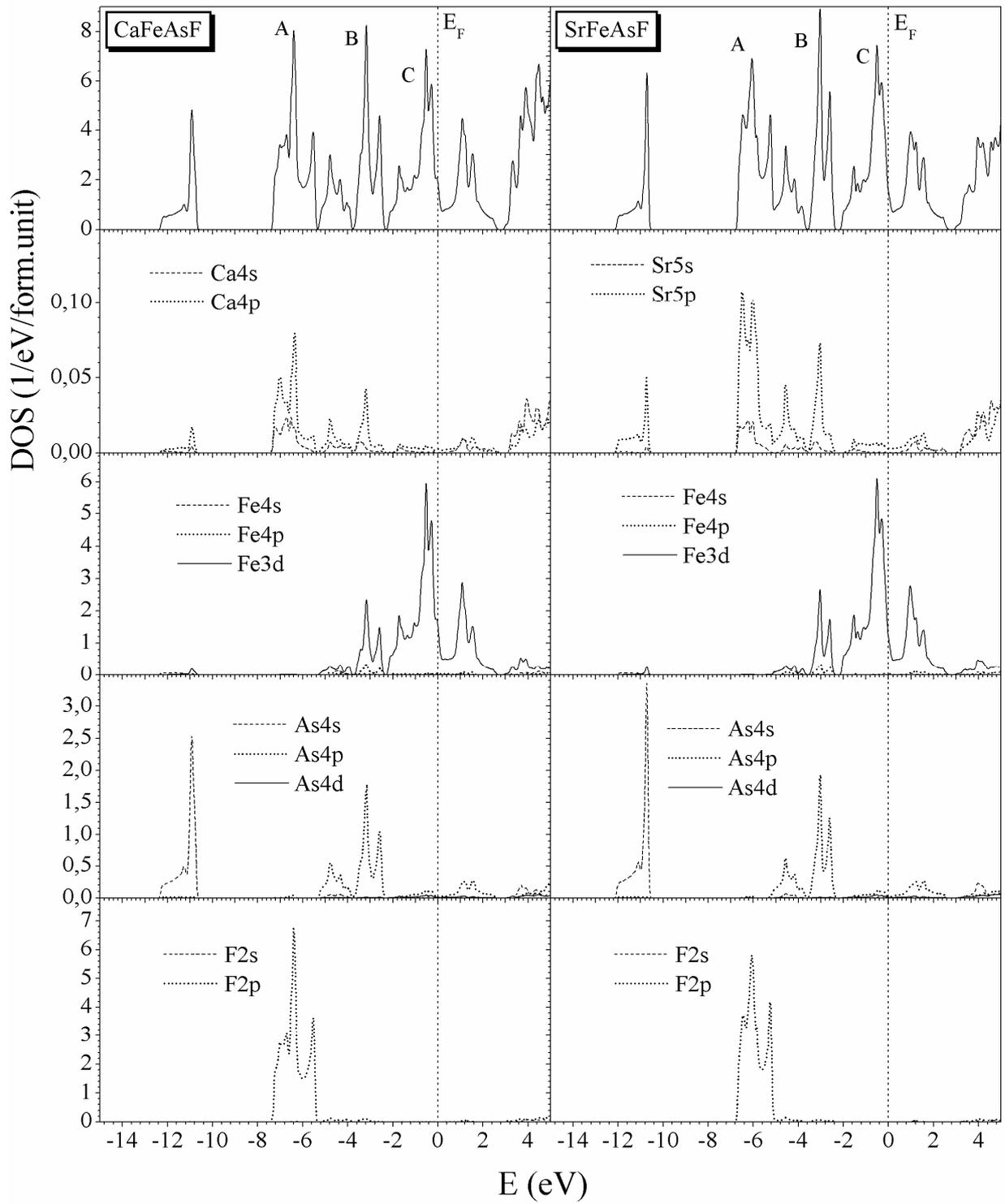

Fig. 2. Total and partial densities of states (DOSs) of CaFeAsF and SrFeAsF.

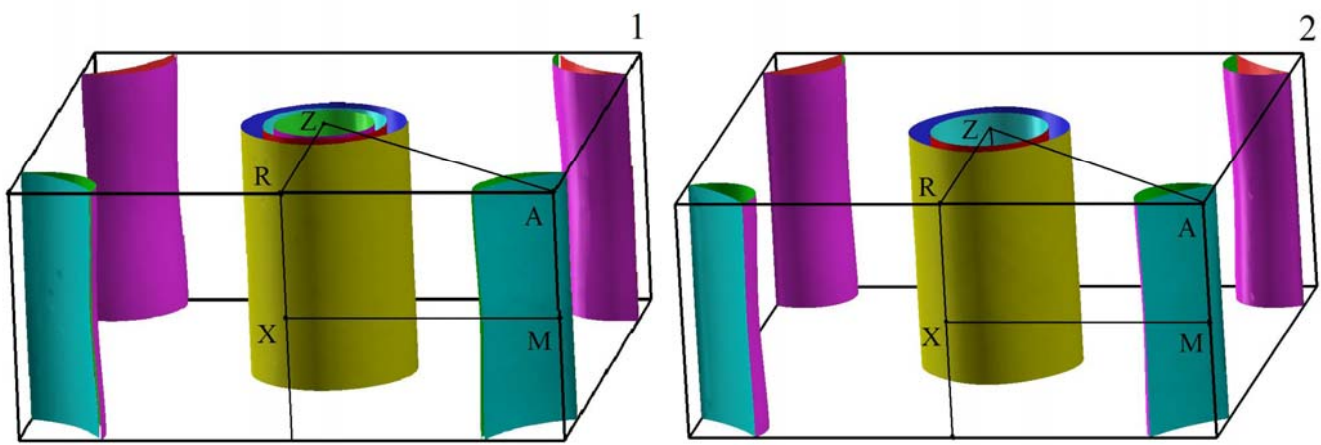

Fig. 3. The Fermi surfaces of CaFeAsF (1) and SrFeAsF (2).